# Effects of a Hovering Unmanned Aerial Vehicle on Urban Soundscapes Perception

Antonio J. Torija[a,b]*, Zhengguang Li[c] and Rod H. Self[a]

[a]ISVR, University of Southampton, Highfield Campus, Southampton, SO17 1BJ, United Kingdom

[b]Acoustics Research Group, Newton Building, University of Salford, Manchester, M5 4WT, United Kingdom

[c]Department of Architecture, Zhejiang University of Science & Technology, Hangzhou, 310023, P.R. China

Author to whom correspondence should be addressed. Electronic mail: A.J.TorijaMartinez@salford.ac.uk

Tel.: +44(0) 0161 295 0469

**Abstract**

Several industry leaders and governmental agencies are currently investigating the use of Unmanned Aerial Vehicles (UAVs), or 'drones' as commonly known, for an ever-growing number of applications from blue light services to parcel delivery. For the specific case of the delivery sector, drones can alleviate road space usage and also lead to reductions in $CO_2$ and air pollution emissions, compared to traditional diesel-powered vehicles. However, due to their unconventional acoustic characteristics and operational manoeuvres, it is uncertain how communities will respond to drone operations. Noise has been suggested as a major barrier to public acceptance of drone operations in urban areas. In this paper, a series of audio-visual scenarios were created to investigate the effects of drone noise on the reported loudness, annoyance and pleasantness of seven different types of urban soundscapes. In soundscapes highly impacted by road traffic noise, the presence of drone noise lead to small changes in the perceived loudness, annoyance and pleasantness. In soundscapes with reduced road traffic noise, the participants reported a significantly higher perceived loudness and annoyance and a lower pleasantness with the presence of the same drone noise. For instance, the reported annoyance increased from 2.3±0.8 (without drone noise) to 6.8±0.3 (with drone noise), in an 11-point scale (0-not at all, 10-extremely). Based on these results, the concentration of drone operations along flight paths through busy roads might aid in the mitigation of the overall community noise impact caused by drones.

# 1.    Introduction

Due to the significant advancement on electrical power, battery and autonomous systems technology, the applications of Unmanned Aerial Vehicles (UAV), or 'drones' as commonly known, seem unlimited (Dorling et al., 2017). An ever-growing number of applications are currently under investigation in sectors such as construction, surveillance and parcel delivery (Yoo et al., 2018). With the continuous increase in consumer demand and cost and time savings in mind, several companies such as Amazon, UPS, Google, and Wal-Mart are testing multi-rotor UAV for delivering small packages or groceries (Alphabet, 2017; BI Intelligence, 2016; Rose, 2013; Vanian, 2017).

The need for reducing greenhouse gas emissions has led to a significant interest in electric propulsion for air vehicles (Schäfer et al., 2019). From the customers' perspective, drone delivery is perceived as more environmentally friendly than delivery by truck, which makes it more appealing for customers who care about the environment (Yoo et al., 2018). Figliozzi (2017) states that UAVs are significantly more efficient for reducing carbon dioxide equivalent emissions than typical diesel delivery vehicles. Several authors suggest that in service zones close to the depot, a deployed UAV based delivery can reduce greenhouse gas and other environmental impacts compared to conventional diesel delivery trucks (Figliozzi, 2017; Goodchild and Toy, 2018; Koitwanit, 2018; Stolaroff et al., 2018).

However, UAV sounds have been found more annoying than sounds of delivery road vehicles (Christian and Cabell, 2017). Although the authors highlighted the uncertainty as to whether the differences in annoyance were due to the particular UAV manoeuvres measured (i.e. farther/slower than for road vehicles measurements) or qualitative differences between UAV and road traffic sounds, Christian and Cabell (2017) found an offset of 5.6 dB between UAV and road vehicles. This means that UAV sounds 5.6 dB lower in A-weighted Sound

Exposure Level (SEL) than road vehicles sounds were reported equally annoying as the latter ones.

The noise generated by UAVs does not qualitatively resemble the noise of conventional aircraft (Cabell et al., 2016; Christian and Cabell, 2017; Torija et al., 2019b; Zawodny et al., 2016); also, compared to contemporary aircraft, UAVs will operate much closer to the public. This is why there is an important uncertainty as to how the public will react to UAV noise. What is clear is that, if not appropriately addressed, noise issues might put at risk the expansion of the UAV sector in urban areas (Theodore, 2018).

This paper is aimed to investigate the noise impact of UAV operations in urban soundscapes. The specific objectives of this research are: (1) Evaluate the impact of the noise generated by the hover of a small quadcopter on the reported loudness, annoyance and pleasantness of different urban soundscapes. (2) Assess the influence of the overall sound level, particular acoustics characteristics of the quadcopter (Cabell et al., 2016; Christian and Cabell, 2017; Torija et al., 2019b; Zawodny et al., 2016) and non-acoustic factors such as visual scene (Liu et al., 2014; Ren and Kang, 2015; Viollon et al., 2002) on the perception of soundscapes with a hovering UAV. (3) Discuss the effect of ambient road traffic noise in masking UAV noise as a potential action for mitigating the noise impact of UAV operations in urban environments.

Aural-visual scenarios were created to investigate the effects of the noise of a small quadcopter hover on the perception of seven urban soundscapes with varying sound level ($L_{Aeq}$), and with varying sound sources. The soundscapes evaluated include sites at varying distances from traffic roads (i.e. 5 m, 50 m and 150 m away) and a park with no influence of road traffic and dominant sounds from birds and a water stream. In order to assess the combined effect of road traffic (at varying levels) and drone noise on soundscape perception, the recordings were carried out in open spaces both alongside a busy traffic junction in city centre and a busy road in the surroundings of the city. The selection of these two areas was to include traffic under typical urban conditions, and also more fluid/high speed traffic. A combination of audio and visual techniques was implemented to create a series of scenarios simulating the operation of a small quadcopter hover in the different urban spaces tested. These audio-visual scenarios provided realistic experiences to the participants of the experiments, allowing more accurate information about the reactions to this novel noise source (Maffei et al., 2013, Ruotolo et al., 2013). The perception of the overall environment is multisensory in its very nature, and both audio and visual factors have been found highly influential in the reported annoyance of transportation systems (Jiang and Kang, 2016; Jiang and Kang, 2017) and wind farms (Schäffer et al., 2019; Szychowska et al., 2018).

This paper is structured as follows: Section 2 explains the acquisition of audio-visual signals, describes the equipment, stimuli and methodology used for the development of experiments, and introduces the data analysis techniques used; In Section 3 and 4 the experimental results are presented and discussed respectively.

## 2. Material and methods

### 2.1. Data collection

The stimuli used in the experiment reported in this paper contain audio and panoramic video signals, which were extracted from a series of indoors and outdoors recordings. Audio-visual recordings were made to capture representative samples of soundscapes with different influence of road traffic noise (see Table 1). Due to the current legislation in the UK[1], forbidding flying drones at least 50 m away from people and property, the audio-visual signals of a small quadcopter were recorded in an anechoic chamber, used for aircraft noise and aeroacoustics research. These audio-visual signals were combined with the audio-visual signals recorded outdoors to generate the stimuli used in the experiment (described below). This approach also allowed the analysis of the effects of exactly the same audio-visual drone stimulus on different urban soundscapes.

### 2.1.1. Outdoors recordings

Fig. 1 shows the (audio-visual) field recording locations in the two areas selected in the city of Southampton (UK).

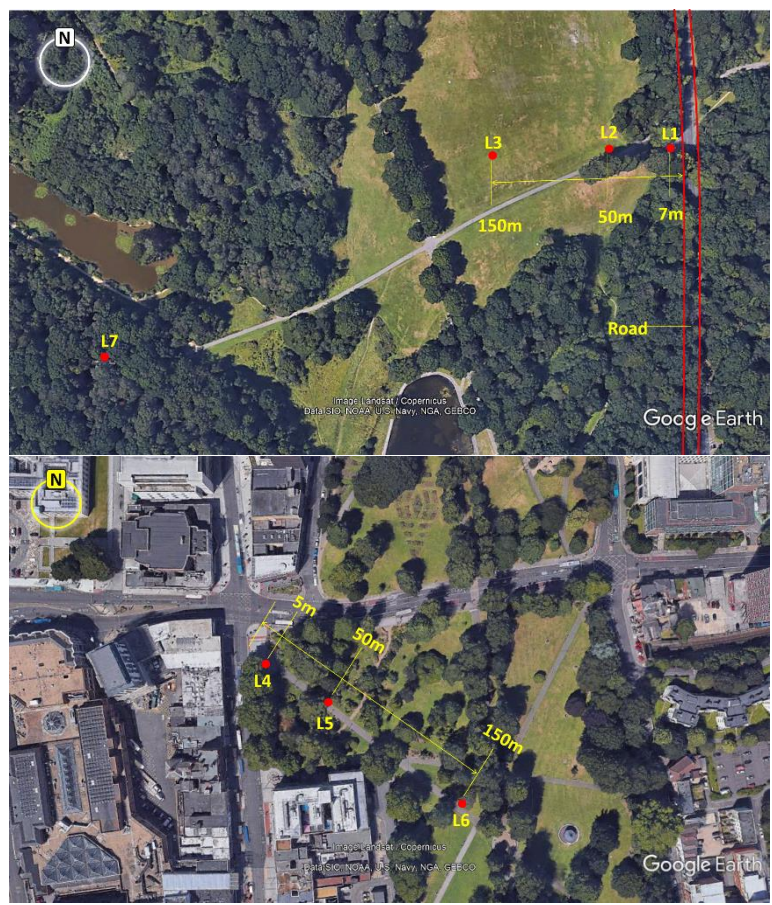

Figure 1. Audio-visual recording sites.

A panoramic camera (Ricoh Theta V) was used to record a high-quality 360° video (30 fps @ 3840 x 1920 pixels or 4K resolution with a data-rate of 56 Mbps; audio bit rate of 96 kbps, audio sample rate of 48.000kHz; MPEG-4 type) in the seven locations selected: 4 in the Common park at varying distances (see Fig. 1) from a busy road with fluid/high speed traffic, and 3 in a park located in the city centre of Southampton (UK) at varying distances (see Fig. 1) from a busy traffic junction (with pulsed-flow traffic conditions typical of urban areas). The audio signals at these locations were recorded via four Micro Electrical-Mechanical System (MEMS) microphones integrated into the panoramic camera to independently record sound from four different directions. These four microphones are arranged as a tetrahedron to get 1st Ambisonic audio in A-format. Then the A-format audio was transferred to B-format using Ricoh Theta software. MEMS are stable and reliable small size microphones with low power consumption. MEMS has an excellent stability across a wide temperature range, and a consistent flat frequency response in the audio frequencies range (especially good at low frequencies) (Lewis and Moss, 2013).

A calibrated class 1 sound meter (Brüel & Kjær 2260 Investigator) was also used to measure the A-weighed sound pressure levels ($L_{Aeq}$) at the site during the recording. The panoramic camera was placed on a tripod at a height of 1.6m from the ground while the sound meter was placed at a height of 1.2m from the ground. Fig. 2 shows a picture of one of the recording sites (location L1).

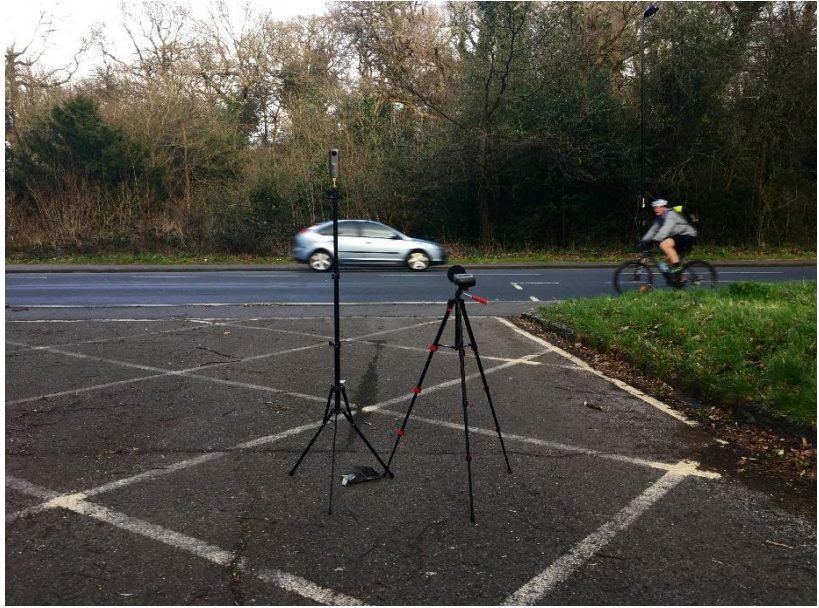

Figure 2. Picture of the recording site in location L1.

### 2.1.2. Anechoic recordings

The recordings of a small quadcopter (DJI Phantom 3 Standard) were carried out in the Anechoic Doak Laboratory at the Institute of Sound and Vibration Research (ISVR). This specific model has a full weight (battery and propellers included) of 1216 g, the max rpm of the propellers is about 7500 and the max load is 2.3 kg (including its own weight). This type of drone is a representative small consumer-level vehicle very promising to be used in construction inspection, surveillance, parcel delivery and traffic control. The quadcopter was fixed to a stand at a distance of 1.8 m above the ground such that only the four rotor blades could move. The same panoramic camera (with a four-channel built-in microphone) used in the recordings outdoors was placed on another tripod at a height of 1.6m from the ground and 0.75 m away from the tripod of the quadcopter. To ease the combination of the panoramic visual signals of the drone and soundscapes recorded, a 3m × 6m green cloth screen was fix behind the quadcopter. To avoid sound reflection effects on the recorded audio signals, a green screen with high acoustic permeability was selected. During the measurements in the anechoic chamber no effect of the green screen was observed in the recorded sound levels. A picture and schematic diagram of the recording setup are shown in Fig. 3. During the recordings, the quadcopter was operated at full power.

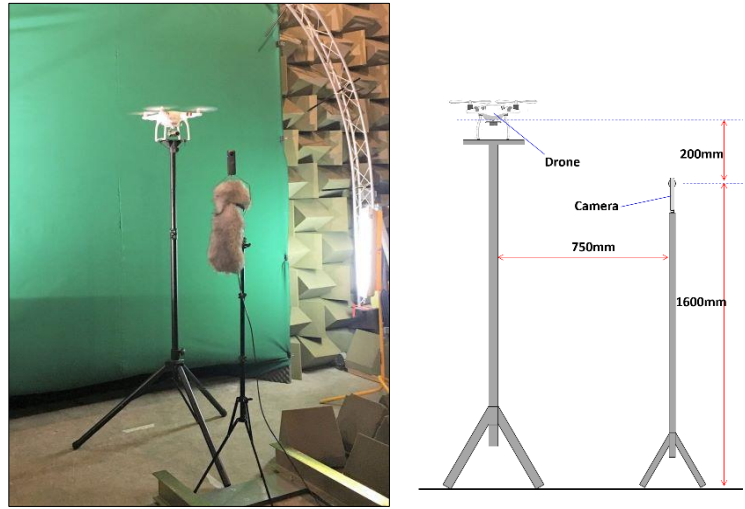

Figure 3. Picture and schematic diagram of the measurement setup at the Anechoic Doak Laboratory at the Institute of Sound and Vibration Research (ISVR).

## 2.2. Stimuli

Two types of stimuli were used in this experiment, i.e. audio only (part 1 of the experiment) and panoramic video with the same audio signals of part 1 (part 3 of the experiment). The results of part 2 are not considered in this paper, as they fall out of its scope (see Section 2.3.3).

### 2.2.1. Processing of the audio signals

A 15 s video excerpt with steady sound level to capture the ambient sound representative of each of the seven locations was selected from the each of the original panoramic video recordings. A 15 s video excerpt of the panoramic video recorded in the anechoic chamber with the drone operating at full power was also selected. The audio signals recorded in the field and in the anechoic chamber were extracted using the FFmpeg Import/Export library of the audio edit software Audacity (v 2.3.0).

One of the objectives of this research is to assess the perception of urban soundscapes with a small drone hover and different road traffic sound levels. The underlying hypothesis is that road traffic noise can mask drone noise, and then mitigate the adverse effects of drone flyovers. The focus of this research is in the differences in the frequency spectra between road traffic and drone noise (see Fig. 14). For the sake of comparison between participants' responses, and in order to find conclusions statistically valid, it was required that all participants received exactly the same sound signal (i.e. sound level, frequency content, etc.) regardless of the movement of their head. For this reason, a monophonic headphone reproduction was preferred to other spatial audio techniques. In the stimuli simulating a drone hover presented to the participants, the small quadcopter is fixed in a steady position, with the other sound sources in the background. Spatial cues increase immersion and plausibility of sound scenes, and so, several spatial audio reproduction techniques have been proposed and tested to be applied in soundscape research (Hong, et al., 2019; Lam, et al., 2019). However, the spatial aspects of soundscapes are not within the scope of this research.

As described above, the four-channel signal was recorded as a 1st order A-Format ambisonic, and then processed to 1st order B-Format. The monophonic signals used in the experiment was the W channel signal, which is a scaled version of the sound pressure at the centre of the microphone array as seen by an omnidirectional pressure microphone.

The sound levels ($L_{Aeq,15s}$) recorded in the field for each 15 s audio except are shown in Table 1. Three $L_{Aeq,15s}$ (i.e. 70, 60 and 55 dBA) were selected both to provide a wide range of sound levels and as representative of the different urban soundscapes recorded. The same sound levels, 70, 60 and 55 dBA, were assigned to the recorded locations with similar distances to road traffic, to investigate whether the different traffic patterns (e.g. urban vs. road traffic) might have effects on the results. Similarly, the location in the park, dominated by water and birds sounds, was set to 55 dBA to investigate the effect of natural sounds vs. distant background road traffic noise. The sound level (i.e. $L_{\text{Aeq,15s}}$) of each 15 s audio except recorded in the field was adjusted in amplitude, using audacity software, to the corresponding target sound levels shown in Table 1 (see $L_{\text{Aeq,15s}}$ (dBA) after adjustment in amplitude row). The sound levels of the 'ambient plus drone' stimuli (see $L_{\text{Aeq,15s}}$ (dBA) after adjustment in amplitude ('ambient plus drone' sounds) row) are the result of the energetic sum of the $L_{\text{Aeq,15s}}$ (dBA) after adjustment in amplitude of each soundscape tested (see $L_{\text{Aeq,15s}}$ (dBA) after adjustment in amplitude row) and the $L_{\text{Aeq,15s}}$ (dBA) after adjustment in amplitude of the drone (i.e. $L_{\text{Aeq,15s}}$ =65 dBA).

The headphone reproduction was calibrated in sound pressure level using an artificial ear (Brüel & Kjær 4153 Artificial Ear) coupled to a class 1 sound level meter (Brüel & Kjær 2260 Investigator), to the corresponding sound levels shown in Table 1 ($L_{\text{Aeq,15s}}$ (dBA) after adjustment in amplitude and $L_{\text{Aeq,15s}}$ (dBA) after adjustment in amplitude ('ambient plus drone' sounds) rows),without altering neither temporal nor spectral characteristics.

**Table 1**

Sound level ($L_{\text{Aeq,15s}}$) for each 15 s audio excerpt.

| Key | L1 | L2 | L3 | L4 | L5 | L6 | L7 | Drone |
|---|---|---|---|---|---|---|---|---|
| $L_{\text{Aeq,15s}}$ (dBA) as recorded in the field | 69.8 | 57.5 | 51.3 | 65.2 | 59.0 | 52.6 | 48.9 | n.a. |
| $L_{\text{Aeq,15s}}$ (dBA) after adjustment in amplitude | 70.0 | 60.0 | 55.0 | 70.0 | 60.0 | 55.0 | 55.0 | 65.0 |
| $L_{\text{Aeq,15s}}$ (dBA) after adjustment in amplitude ('ambient plus drone' sounds) | 71.2 | 66.2 | 65.4 | 71.2 | 66.2 | 65.4 | 65.4 | n.a. |

The sound level ($L_{Aeq,15s}$) of the quadcopter was set at 65 dBA. This sound level was chosen on the basis of the results of a measurement campaign carried out by Cabell et al (2016) for a series of small quadcopters and hexacopters. Cabell et al (2016) found the sound level of small quadcopters at 15 m from the microphone ranging between 65 and 70 dBA. In the research presented in this paper it was assumed that a hovering altitude of 15-20 m is reasonable, and therefore, 65 dBA was selected as a representative sound exposure to a small quadcopter.

The 'ambient plus drone' audio signals were created by combining with audacity software each of the seven field recorded 15 s excerpt and the 15 s drone audio signal recorded in the anechoic chamber. This resulted in fourteen audio signals (seven with 'ambient' sounds and seven with 'ambient plus drone' sounds) as the stimuli for this experiment.

### 2.2.2. Processing of the panoramic video signals

A series of panoramic videos simulating representative scenarios of all the seven urban soundscapes recorded were used as stimuli in the experiment. Altogether, 14 scenarios were assessed by the participants: the seven original urban soundscapes recorded, and the same seven urban soundscapes with the addition of a small quadcopter hover. The panoramic video of the quadcopter recorded in the anechoic chamber, with green screen background, was keyed out and added onto each of the seven recorded urban soundscapes using a video effects software, i.e. Adobe After Effect CC 2017. In this step, the videos were muted and the corresponding calibrated audio signals (see Section 2.2.1) were imported (see Fig. 4). Therefore, exactly the same sounds were presented to the participants in parts 1 and 3 of the experiment. Before the experiments, the experimenters checked that the reproduced levels in parts 1 and 3 were identical using an artificial ear coupled to a class 1 sound level meter (see Section 2.2.1).

Fig. 5 displays a picture of the viewer's perspective for one of the locations tested (location L4), without and with the drone hover. In each of the seven panoramic videos produced for the 'ambient plus drone' scenarios, the drone was simulated in a fixed position (i.e. hover) showing fully operational propellers rotating at full power (see above max rpm).

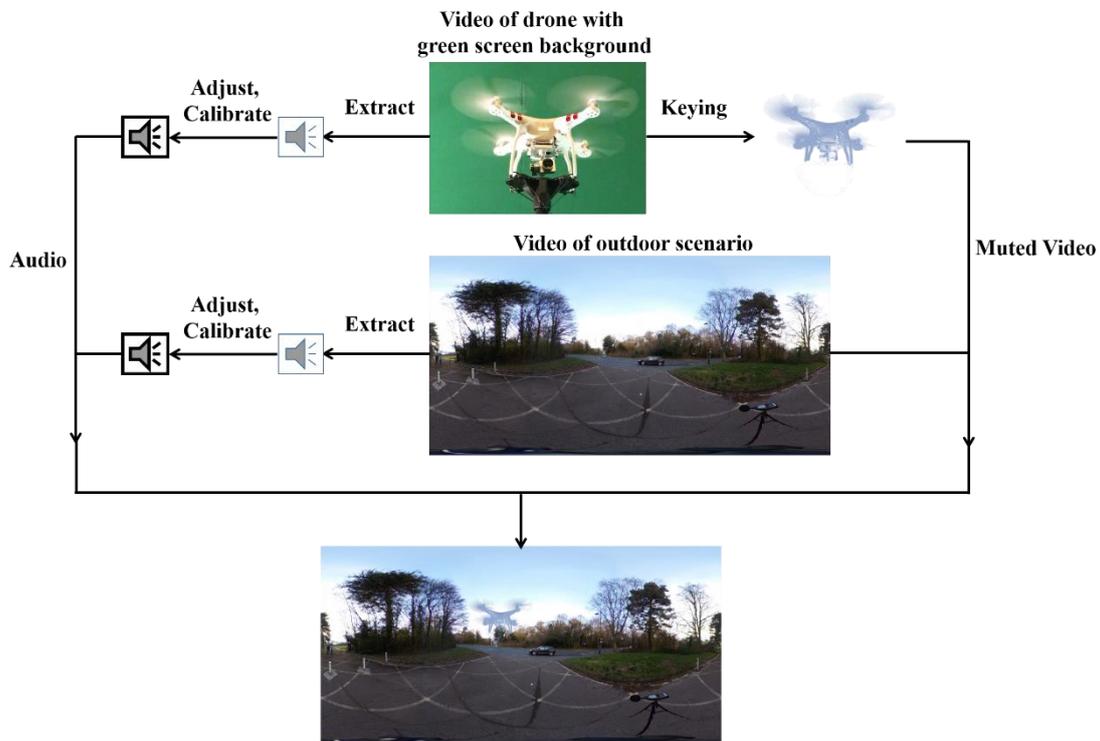

Figure 4. Overview of the processing to create the audio-visual stimuli with the quadcopter hover.

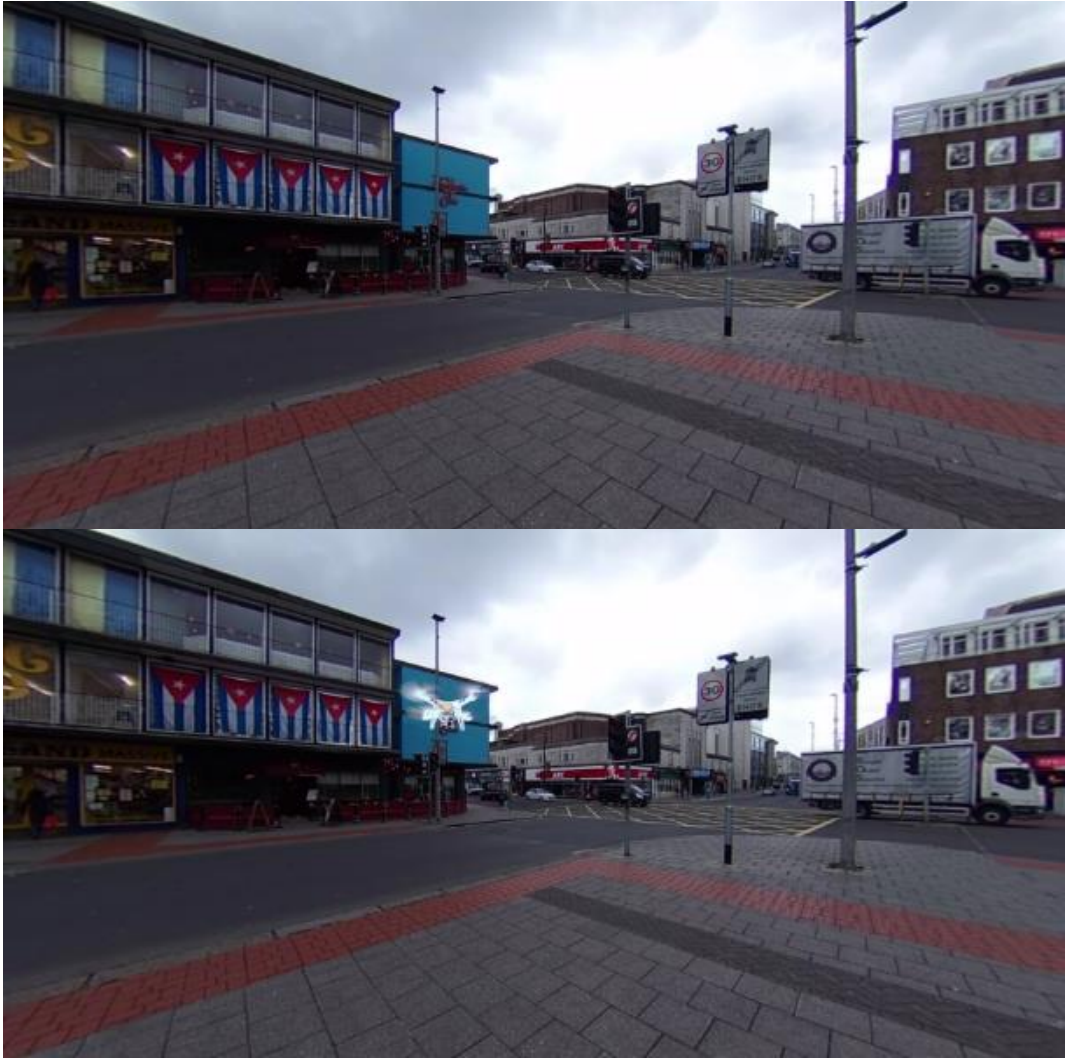

Figure 5. Viewer's perspective for the location L4, without (top) and with the quadcopter hover (bottom).

## 2.3. Listening experiments

### 2.3.1. Participants

The listening tests were undertaken by 30 healthy participants (16 males and 14 females) aged between 21 and 59 years old (mean age = 30.5, standard deviation = 9.2, 57% between 20 and 29 years old, 31% between 30 and 39 years old, 6% between 40 and 49 years old, and 6% between 50 and 59 years old) who were recruited by email within university. A thank you gift of £10 for taking part was used to incentivize participation in the listening tests. Prior to participating in the listening test, each participant was required to confirm normal hearing ability and asked to fill out a consent form. This experiment was approved by the Ethics and Research committee of the University of Southampton.

### 2.3.2. Equipment for the presentation of stimuli

The hardware setup used for the experiments consisted of a powerful desktop computer (Intel Core i7-2600 CPU @3.40GHz, 16.0 GB RAM, 64-bit Windows 10 Operating System) with a high-performance graphics card (NVIDIA GeForce GTX 1080), a USB DAC/headphone amplifier (Audioquest, DragonFly Red v1.2), a pair of open back headphones (AKG K-501), and a Facebook Oculus Rift S virtual reality head-mounted-display (VR HMD).

The order of play was generated by the experimenters before each experiment using a random order generator software (i.e. The Hat Deluxe) to eliminate memory bias from prior judgments. In the first part, the audio stimuli were presented by the experimenter using the media player software VLC media player v3.0.6. In the third part, the participants were instructed to play back themselves the panoramic audio-visual stimuli using the VR video player DeoVR Video Player v5.8. Note that, as mentioned above, the second part of the experiments is not included in this paper. The volume level control on the desktop was blocked, so the reproduced sound levels were not altered after calibration. The tests were carried out in a very quiet environment (i.e. a small anechoic chamber at ISVR), with no interference from outside in order to avoid distractions. The background sound level in this small anechoic chamber was 15.1 dBA.

### 2.3.3. Experimental procedure

This paper reports the results of two out of three parts of a listening experiment. As described above, in the first and third parts of the experiment, only audio signals and audio-visual signals respectively simulating a drone hover in seven urban scenes were presented to the participants. In the second part of the experiment, a series of drone, road vehicles and aircraft sounds were played back, and the participants were requested to rank them by order of preference using a methodology developed by Torija et al. (2019a). The objective of this second part (of 40-min duration) was to compare subjective perception of drone flyovers with aircraft flyovers and road vehicles pass-byes. The data gathered in this second part are not included in the paper, as it falls out of its scope.

The experiments involved a series of assessment tasks, where the participants reported their perception of loudness, annoyance and pleasantness induced by the sounds they heard (first part) or the panoramic videos they heard and watched (third part), using an 11-point scale (0-not at all, 10-extremely). In each part, i.e. only audio and audio plus panoramic video, 14 15-second stimuli were rated, with a 20-second break in between.

Panoramic video recordings and VR HMD were the stimuli and equipment chosen to present the participants with the different scenarios to be evaluated. A VR HMD provides important operational benefits compared to other reproduction equipment, such as big screens. Further, a panoramic video recording enables a better representation and simulation of the locations under study. The use of both panoramic video recordings and VR HMD made the participants more intuitively and better understand the scenarios presented.

For the sake of comparison and statistical validity, all the participants were advised to look at front in order to focus on the area where the drone hover was simulated. During the 20-second break the participants reported their answers, and then rested and waited for the next stimulus. The stimuli were presented (and rated) only once, in a random order. Before the start of the first part of the experiment, several audio samples were presented to the participants; similarly, before the start of the third part, several audio-visual samples were presented to the participants. The objective was to make the participants familiar with the tasks requested during the experiment (including the subjective ratings), and also with the equipment used.

Specifically, audio samples of different loudness were used to instruct the participants in the rating using the 11-point scale, and panoramic video samples were used for the participants to learn how to use the VR video player. After the completion of the experiment, in an informal chat, the participants were inquired as to their views on both the experimental design and the audio/audio plus visual stimuli they heard/heard and watched.

In the first part, the participants reported their responses in a paper questionnaire provided. In the third part, as the participants were wearing the VR HMD, they reported orally their rates after each stimulus, and it was the experimenter who wrote down their answers in a paper questionnaire.

Considering the training/introduction, experiment and debrief, the duration of each part 1 and 3 was 20 min. Altogether, including the three parts of the experiment (second one not reported in this paper), the average total duration of the experiment was 1 hour and 20 min.

## 2.4. Data Analysis

The analysis of the influence of the overall sound level, particular acoustics characteristics of the quadcopter and non-acoustic factors such as visual scene on soundscape perception was addressed using multilevel modelling. Multilevel linear models (also known as mixed models) are a suitable approach to take into account individual responses of participants, as it is assumed that regression parameters (i.e. intercept and slopes) vary randomly across participants (Hox, 2010). As every participant might have a different interpretation of the rating scale, leading to different regression parameters, multilevel linear modelling was assumed an accurate approach to investigate the contribution of each acoustic and non-acoustic factors to the perception of the soundscapes tested. All the statistical analyses were carried out with the statistical package IBM SPSS Statistics 25.

## 3.    Results

### 3.1.    Perception of urban soundscapes with a hovering drone

Fig. 6 shows the perceived loudness reported by the participants of the listening experiments for the seven urban locations tested, with and without the presence of the noise generated by a small quadcopter hover (e.g. L1 vs. L1D), also differentiating between the cases with and without visual stimuli.  In locations L1 and L4, the closest to road traffic, the presence of drone noise has a limited effect with an increase in reported loudness of 9% and 15% (L4 and L1 respectively).  As the distance from the road traffic increases, and therefore the ambient sound level decreases, the effect of drone noise in reported loudness also increases, from 46% in L5 to 99% in L3.  The highest increase in reported loudness is observed in location L7 (park with water and birds sounds), where the reported loudness with drone noise is 2.2 times the one reported for the typical ambient sound. The visual stimuli seem not to have a clear effect on the reported loudness. In locations with high ambient sound levels, i.e. L1 and L4, the reported loudness decreases with visual stimuli. However, in the locations with low ambient sound levels, the reported loudness is slightly higher with visual stimuli.

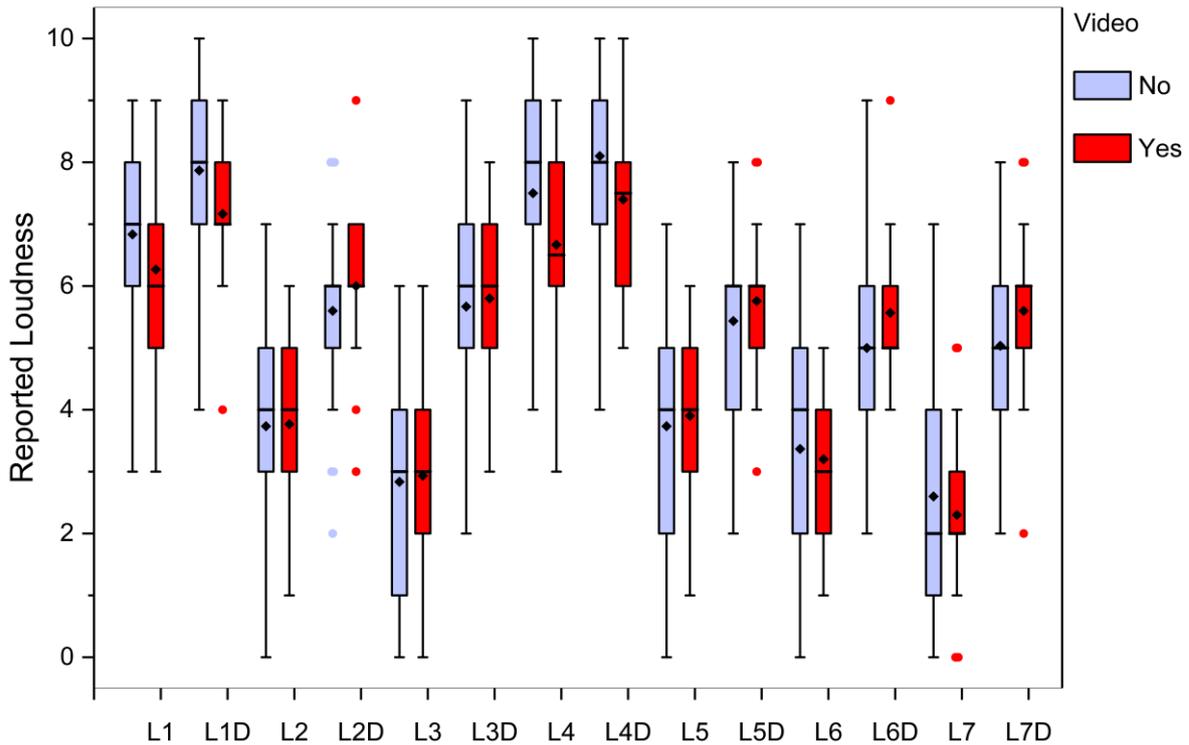

Figure 6. Reported loudness in each of the seven urban soundscapes evaluated without and with the noise generated by the drone hover (e.g. L1 vs. L1D), and without and with panoramic video.

In Fig. 7, it is shown the reported annoyance for the seven urban locations tested for the conditions with and without noise of a small quadcopter hover, and with and without visual stimuli. The reported annoyance increases between 24% and 28% (locations L4 and L1 respectively) with the presence of drone noise in locations with high ambient road traffic noise. In locations with little influence of road traffic noise, and consequently low ambient sound levels, significant increases in the reported annoyance are observed with the presence of drone noise. In these locations the increase in reported annoyance with drone noise ranges between 2.3 (locations L2 and L5) and 6.3 (location L7) times the reported annoyance for ambient noise. In fact, the median value of the reported annoyance in all the urban locations tested was about 7 (in a 11-point scale from 0 to 10) with drone noise, regardless the overall sound levels.

Comparing the responses with and without visual stimuli, the reported annoyance is slightly lower with visual stimuli in all the urban locations (8% lower than without visual stimuli).

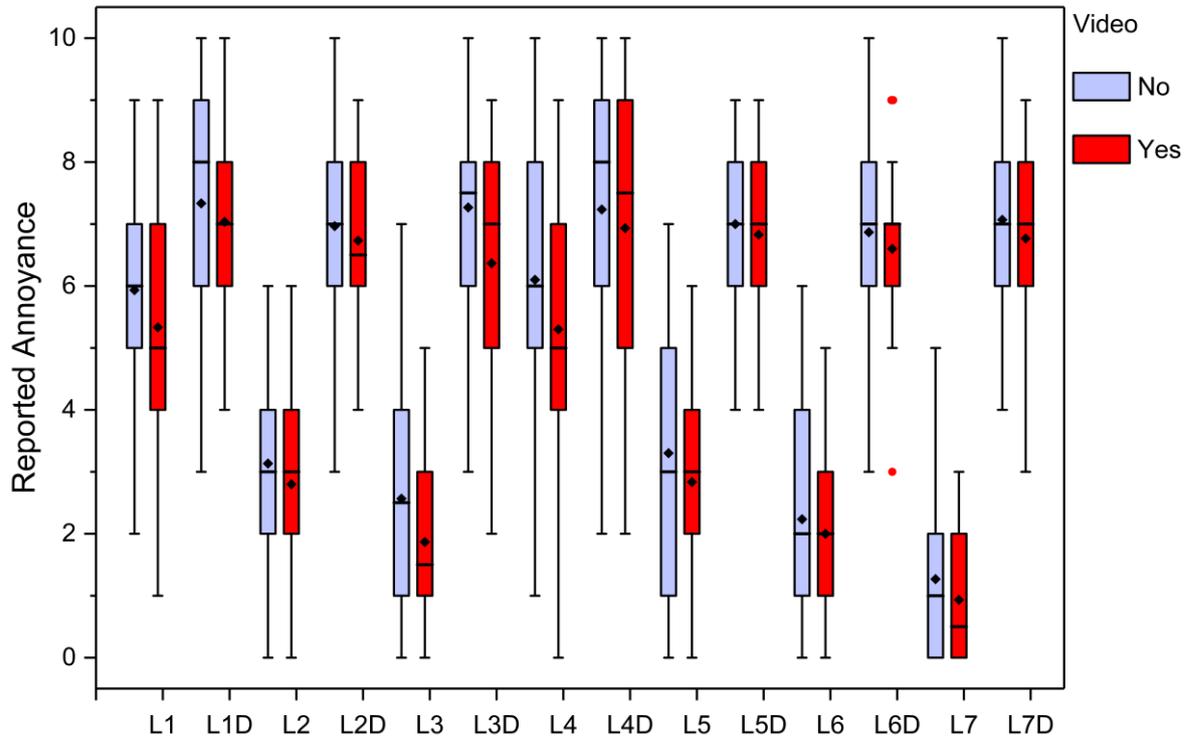

Figure 7. Reported annoyance in each of the seven urban soundscapes evaluated without and with the noise generated by the drone hover (e.g. L1 vs. L1D), and without and with panoramic video.

Fig. 8 shows the reported pleasantness for the seven urban locations tested with and without noise generated by a small quadcopter hover, and also with and without visual stimuli. The reported pleasantness, with and without drone noise, in locations with high road traffic noise is similar, i.e. median = 0.8 and 1.5 with and without drone noise respectively. In locations with reduced influence of road traffic noise, and also water and birds sounds (location L7), the reported pleasantness without drone noise is significantly higher than with drone noise. In these locations, the reported pleasantness without drone noise is from 2.9 (location L5) to 4.0 (location L7) times higher than with drone noise. The influence of the visual stimuli is observed to have a larger influence than in the previous two cases (i.e. reported loudness and annoyance). Comparing the responses with and without visual stimuli, the reported pleasantness is notably higher with visual stimuli in all the urban locations (47% higher than without visual stimuli).

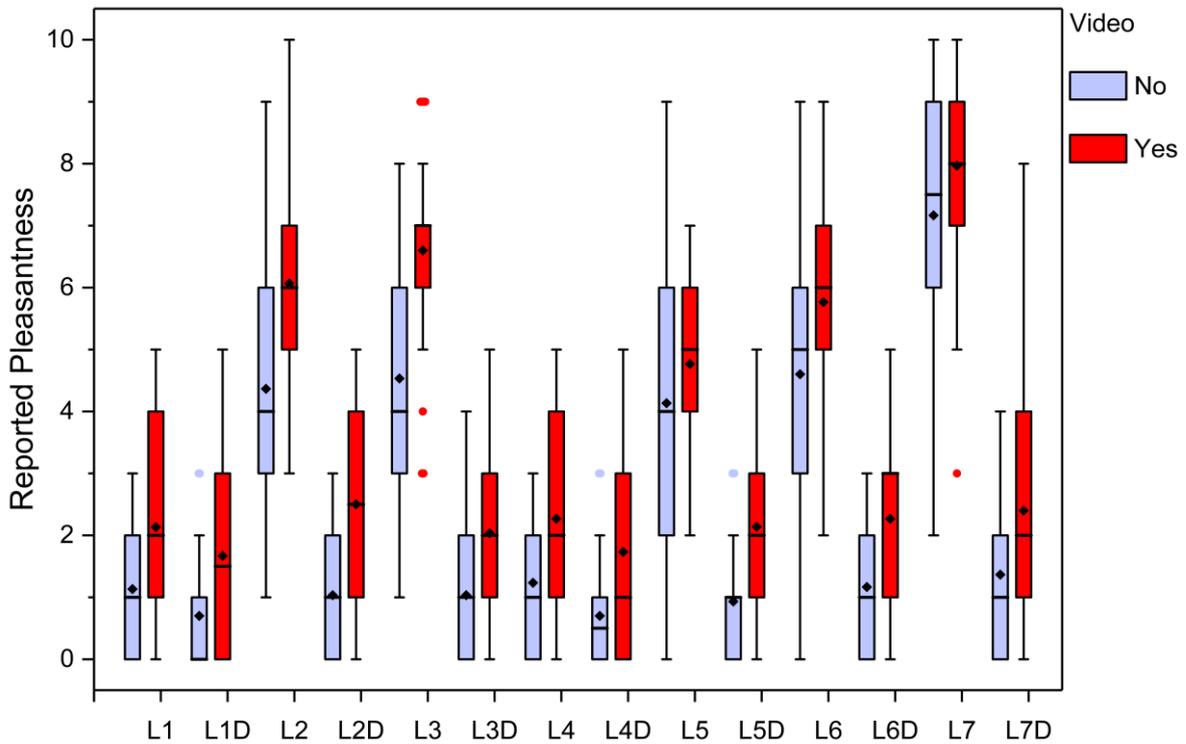

Figure 8. Reported pleasantness in each of the seven urban soundscapes evaluated without and with the noise generated by the drone hover (e.g. L1 vs. L1D), and without and with panoramic video.

**Table 2**

Results of the related-samples Friedman's two-way analysis of variance by ranks. It is shown the pairwise comparisons with statistically significant differences (p<0.05) between the conditions: C1 ('ambient', 'only audio'), C2 ('ambient plus drone', 'only audio'), C3 ('ambient', 'audio plus video') and C4 ('ambient plus drone', 'audio plus video').

| L1 | | | |
|---|---|---|---|
| Pairwise Comparisons | Reported Loudness | Reported Annoyance | Reported Pleasantness |
| C1-C2 | p<0.05 | p<0.05 | |
| C1-C3 | | | p<0.05 |
| C2-C4 | | | p<0.05 |
| C3-C4 | | p<0.05 | |
| L2 | | | |
| Pairwise Comparisons | Reported Loudness | Reported Annoyance | Reported Pleasantness |
| C1-C2 | p<0.05 | p<0.05 | p<0.05 |
| C1-C3 | | | |
| C2-C4 | | | p<0.05 |
| C3-C4 | p<0.05 | p<0.05 | p<0.05 |
| L3 | | | |
| Pairwise Comparisons | Reported Loudness | Reported Annoyance | Reported Pleasantness |
| C1-C2 | p<0.05 | p<0.05 | p<0.05 |
| C1-C3 | | | |
| C2-C4 | | | |
| C3-C4 | p<0.05 | p<0.05 | p<0.05 |
| L4 | | | |
| Pairwise Comparisons | Reported Loudness | Reported Annoyance | Reported Pleasantness |
| C1-C2 | | | |
| C1-C3 | | | p<0.05 |
| C2-C4 | | | p<0.05 |
| C3-C4 | | p<0.05 | |
| L5 | | | |
| Pairwise Comparisons | Reported Loudness | Reported Annoyance | Reported Pleasantness |
| C1-C2 | p<0.05 | p<0.05 | p<0.05 |
| C1-C3 | | | |
| C2-C4 | | | |
| C3-C4 | p<0.05 | p<0.05 | p<0.05 |
| L6 | | | |
| Pairwise Comparisons | Reported Loudness | Reported Annoyance | Reported Pleasantness |
| C1-C2 | p<0.05 | p<0.05 | p<0.05 |
| C1-C3 | | | |
| C2-C4 | | | |
| C3-C4 | p<0.05 | p<0.05 | p<0.05 |
| L7 | | | |
| Pairwise Comparisons | Reported Loudness | Reported Annoyance | Reported Pleasantness |
| C1-C2 | p<0.05 | p<0.05 | p<0.05 |
| C1-C3 | | | |
| C2-C4 | | | |
| C3-C4 | p<0.05 | p<0.05 | p<0.05 |

A Friedman's two-way analysis of variance by ranks was conducted to investigate whether there are statistically significant differences, in the responses of the participants about perceived loudness, annoyance and pleasantness, between four conditions: C1 ('ambient', 'only audio)', C2 ('ambient plus drone', 'only audio'), C3 ('ambient', 'audio plus video') and C4 ('ambient plus drone', 'audio plus video'). As shown in Table 2, in locations with little influence of road traffic noise (i.e. L2, L3, L5, L6 and L7) there are statistically significant differences (p<0.05) in the reported loudness, annoyance and pleasantness between the conditions 'with drone and 'without drone' noise, both without and with visual stimuli. In location L1 (by the side of a busy road), statistically significant differences in the reported loudness and annoyance are observed between the conditions 'with drone' and 'without drone' noise, with only audio stimuli; and statistically significant differences in the reported annoyance between the conditions 'with drone' and 'without drone' noise, with audio plus visual stimuli. In location L4 (by the side of a street with busy traffic), statistically significant differences in the reported annoyance are observed between the conditions 'with drone' and 'without drone' noise, with audio plus visual stimuli. In locations L1 and L4, statistically significant differences in the reported pleasantness are also observed between the conditions 'only audio stimuli' and 'audio plus visual stimuli', both with only 'ambient' noise and with 'ambient plus drone' noise. As described above, in these locations, the perceived pleasantness reported by the participants with visual stimuli is notably higher than with only audio stimuli.

## 3.2. Relationship between $L_{\text{Aeq}}$ and subjective ratings for urban soundscapes with a drone hover

The sound levels ($L_{\text{Aeq}}$) set for each of the seven urban location tested, with and without drone noise (14 scenarios in total), range from 55 dBA to 71.2 dBA (see Table 1). The relationship between $L_{\text{Aeq}}$ and reported loudness, annoyance and pleasantness for the whole set of urban soundscape scenarios evaluated is shown in Figs. 9 and 10. The values of reported loudness, annoyance and pleasantness displayed in Figs. 9 and 10 for each scenario evaluated correspond to the median value calculated from all participants' responses.

Fig. 9 shows the relationship between $L_{Aeq}$ and reported loudness (top), annoyance (middle) and pleasantness (bottom) for the conditions 'only audio' (circles) and 'audio plus video' (triangles). As observed in Fig. 9 – top, the slope (i.e. s = Δ subjective rating / Δ $L_{Aeq}$) in the relationship $L_{Aeq}$ vs. reported loudness is similar for both condition 'only audio stimuli' (s = 0.30) and condition 'audio plus visual stimuli' (s = 0.27). For the relationship $L_{Aeq}$ vs. reported annoyance (Fig. 9 – middle), the slopes of both conditions (i.e. 'only audio' and 'audio plus video') are almost the same (s = 0.37 and 0.35). However, in this case an offset of 1.2 dB is observed between both conditions, i.e. for a given value of reported annoyance, the $L_{Aeq}$ of the condition 'audio plus visual stimuli' is 1.2 dB higher than for the condition 'only audio stimuli'. For the relationship $L_{Aeq}$ vs. reported pleasantness (Fig. 9 – bottom), the slope is similar for both condition 'only audio stimuli' (s = -0.34) and condition 'audio plus visual stimuli' (s = -0.38). An offset of 3.9 dB is observed between both conditions, i.e. for a given value of reported pleasantness, the $L_{Aeq}$ of the condition 'audio plus visual stimuli' is 3.9 dB higher than for the condition audio stimuli. This significant offset seems to indicate (as described above in Section 3.1) that the visual stimuli influence the perceived pleasantness.

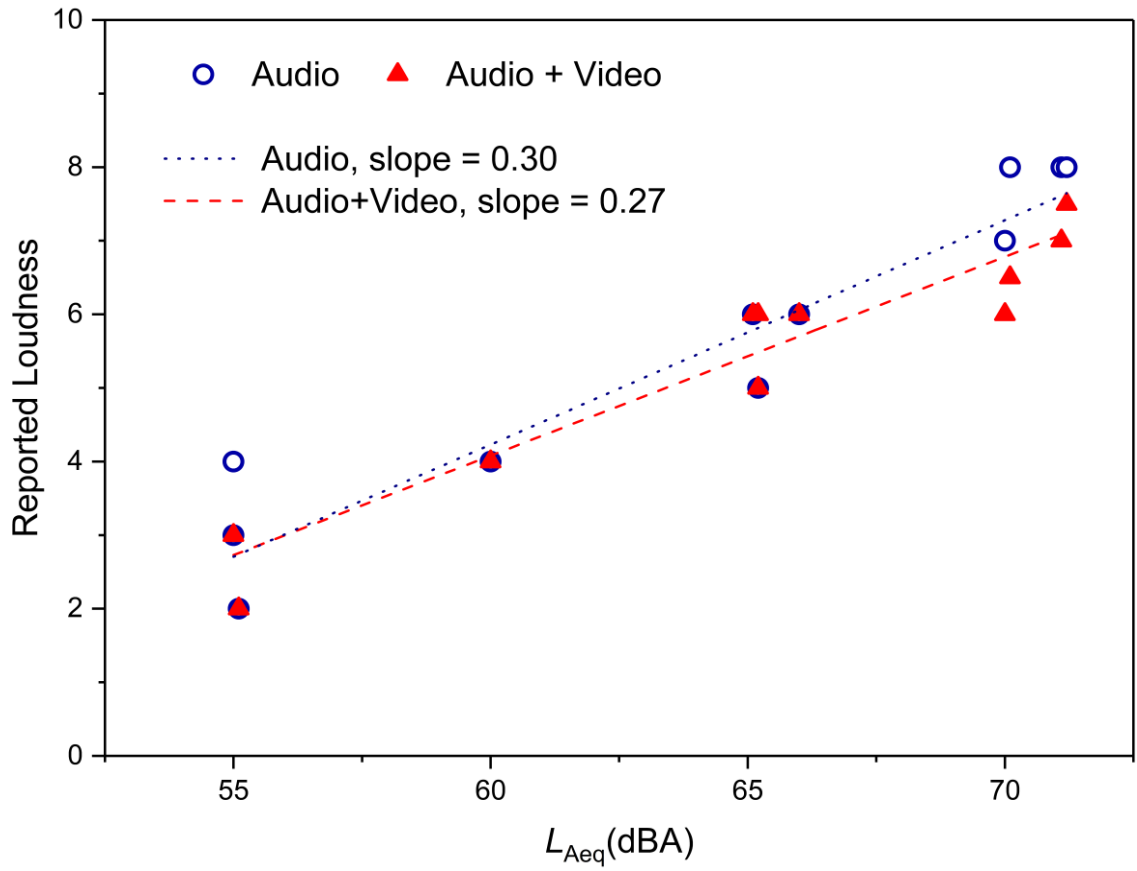

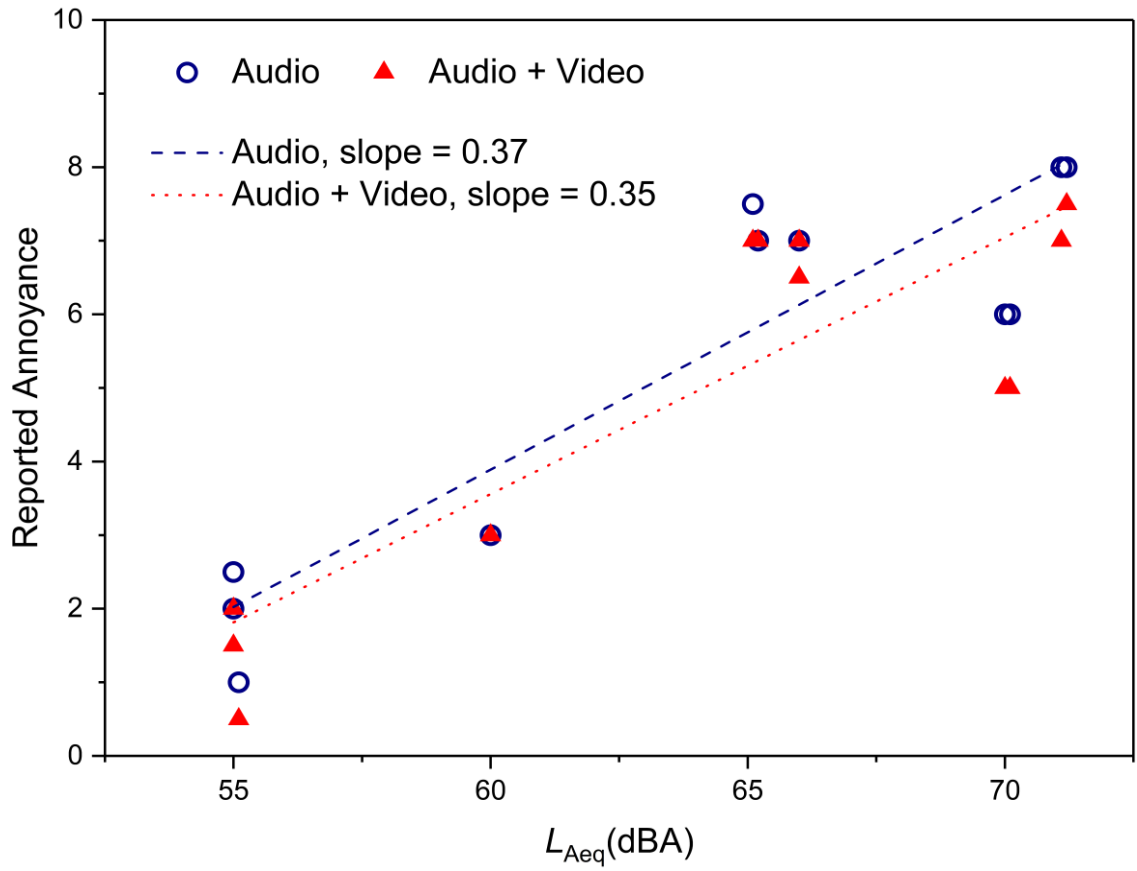

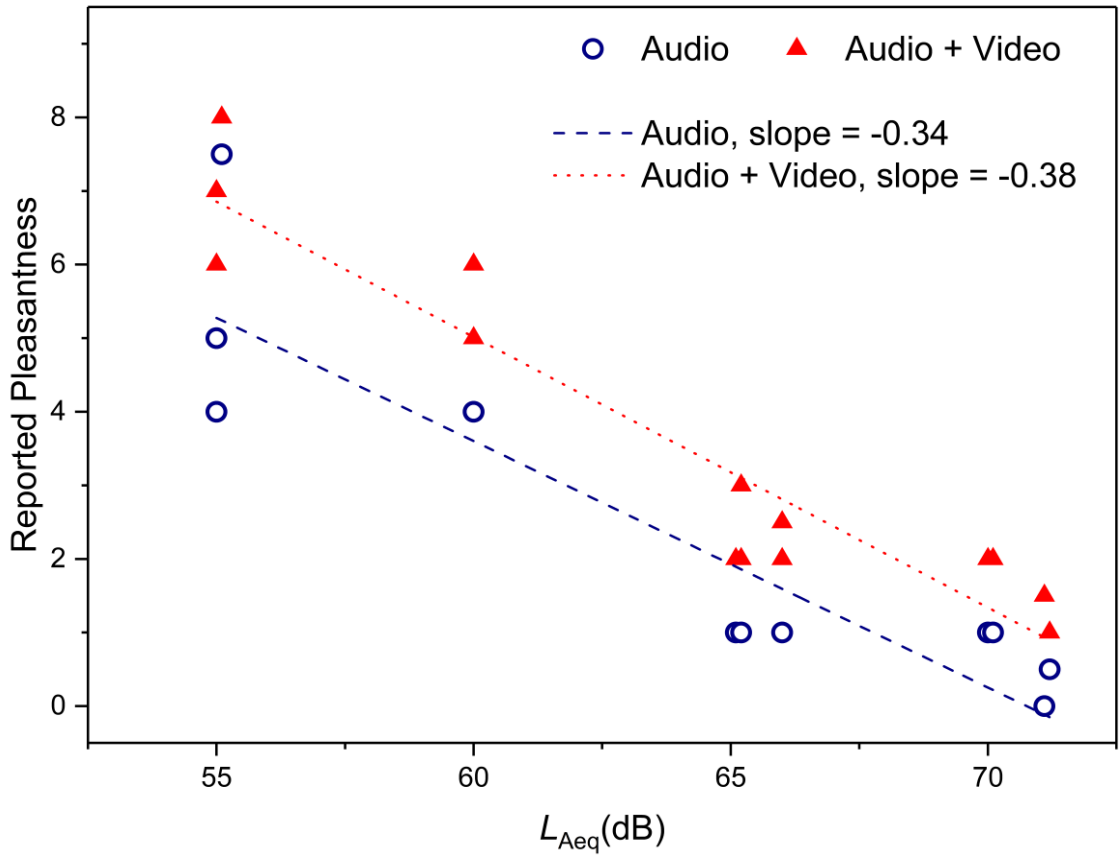

Figure 9. $L_{Aeq}$ vs. reported loudness (top), annoyance (middle) and pleasantness (bottom) for the conditions 'only audio' (circles) and 'audio plus video' (triangles).

The relationship between $L_{Aeq}$ and reported loudness (top), annoyance (middle) and pleasantness (bottom) for the conditions 'ambient' (triangles) and 'ambient plus drone' (circles) is shown in Fig. 10. Fig. 10 – top, i.e. relationship between $L_{Aeq}$ vs. reported loudness, shows that the slope for the condition 'ambient plus drone' is higher (s = 0.34) than for the condition 'ambient' (i.e. without drone) (s = 0.27). For both conditions, the responses on perceived loudness seem mainly driven by $L_{Aeq}$. The relationship between $L_{Aeq}$ vs. reported annoyance (Fig. 10 – middle), seems mainly driven by $L_{Aeq}$ for the condition 'ambient' (s = 0.26). However, for the condition 'ambient plus drone', the reported annoyance is about 7 in all locations regardless of the $L_{Aeq}$. If we assume that the relationship between annoyance and $L_{Aeq}$ is approximately linear in the sound level range between 50 dBA and 75 dBA, the difference between two curves at the 65 dBA reach about 2 units, yielding a difference of 6 dB equivalent. This suggests that the participants' responses on perceived annoyance are highly influenced by acoustics factors, other than sound level, particularly characteristic of small quadcopter noise (Cabell et al., 2016; Christian and Cabell, 2017; Torija et al., 2019b; Zawodny et al., 2016), or non-acoustics factors such as visual scene (Jiang and Kang, 2016; Jiang and Kang, 2017; Schäffer et al., 2019; Szychowska et al., 2018) and expectation (Bruce and Davies, 2014; Perez-Martinez et al., 2018). Fig. 10 – bottom shows that the relationship between $L_{Aeq}$ vs. reported pleasantness seems also driven by $L_{Aeq}$ for the condition 'ambient' (s = -0.32). As for the case of reported annoyance, the participants' responses on perceived pleasantness for the condition 'ambient with drone' seems highly influenced by acoustics or non-acoustics factors associated to drone noise. In Fig. 10 – bottom, it is also observed a higher degree of variability in the responses on perceived pleasantness, which might be due to the effect of visual stimuli on the reported pleasantness, as described above (Section 3.1).

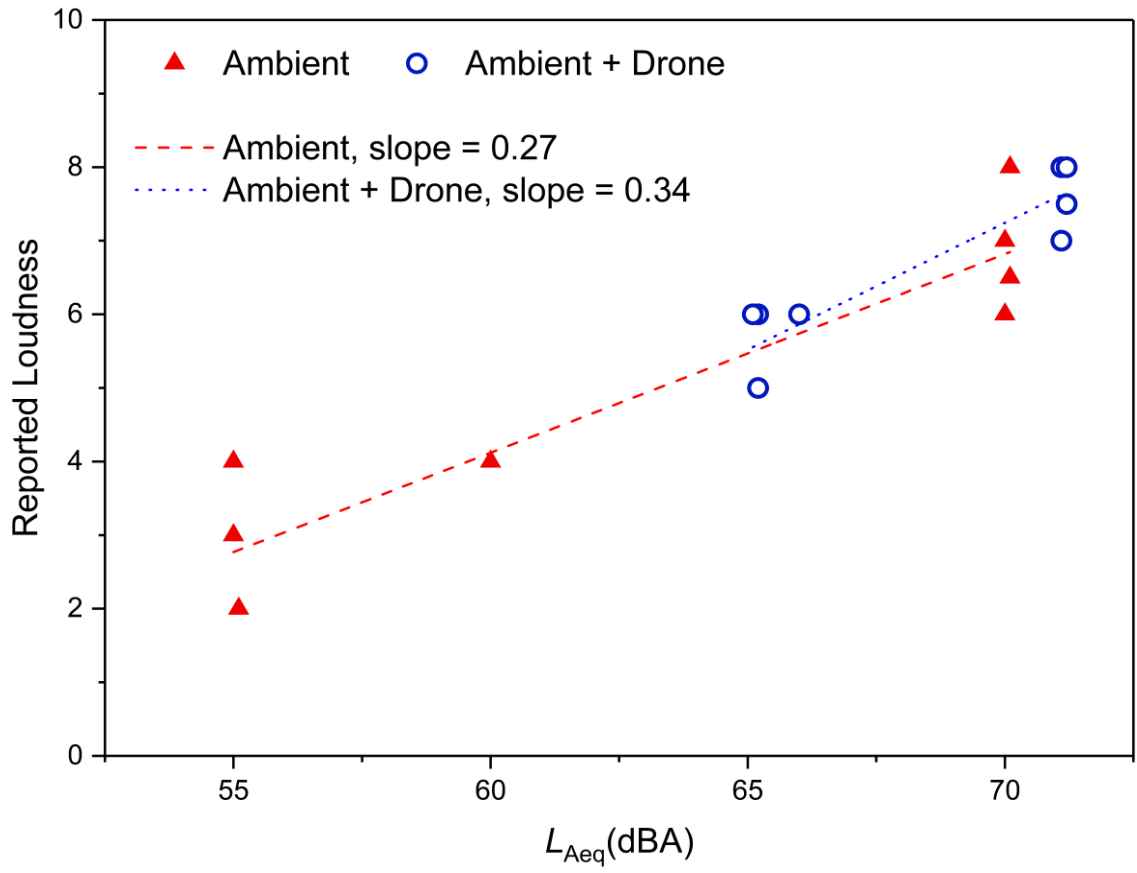

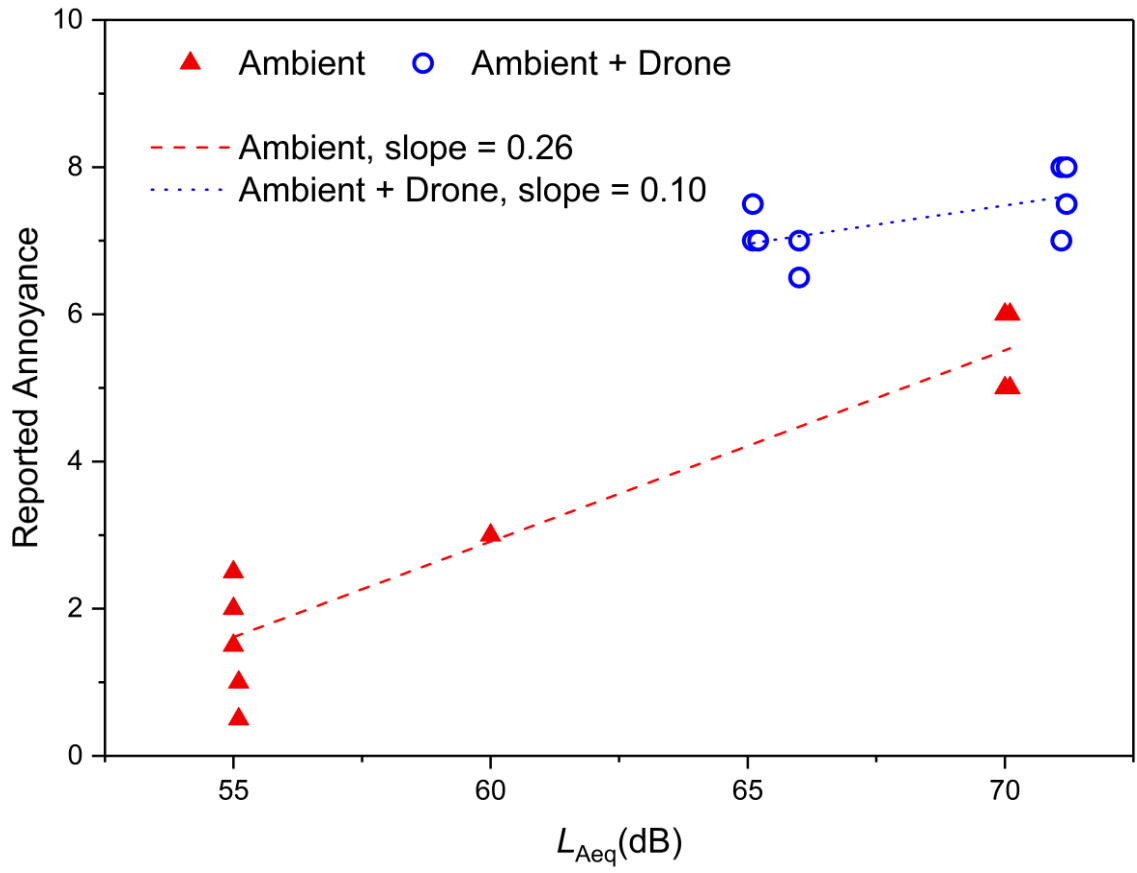

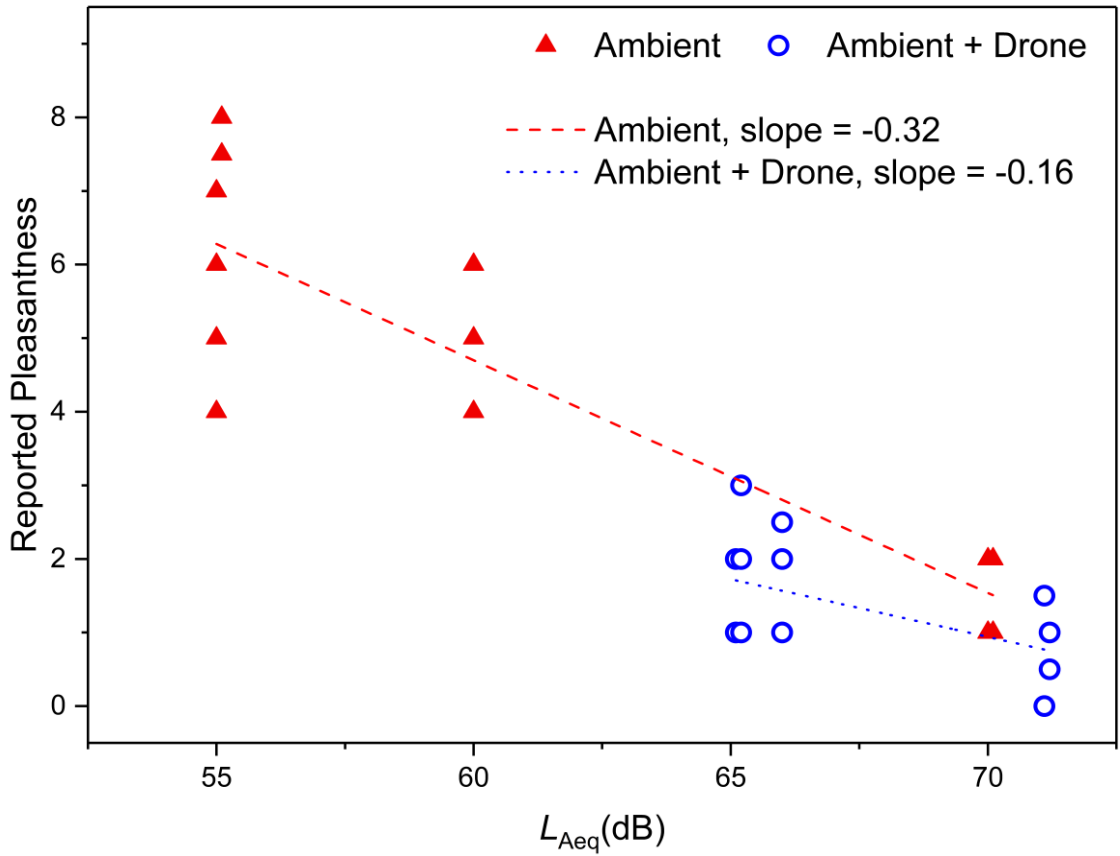

Figure 10. $L_{Aeq}$ vs. reported loudness (top), annoyance (middle) and pleasantness (bottom) for the conditions 'ambient' (triangles) and 'ambient plus drone' (circles).

## 3.3. Importance of acoustics and non-acoustics factors of drone noise on urban soundscapes perception

The importance of each factor, i.e. $L_{Aeq}$, drone noise source and visual scene, on the reported loudness, annoyance and pleasantness was evaluated using a "one-off" approach. In this approach, the importance of each factor is assessed based on model accuracy when removing it from the analysis (Boucher et al., 2019). Three multilevel linear regression models were tested, M1 (fixed intercept, fixed slopes), M2 (fixed intercept, variable slopes) and M3 (variable intercept, variable slopes). The variable parameters in models M2 and M3 represent random effects. Based on models' results, it is first observed that participant is a significant factor, and after participant is taken into account, reported loudness, annoyance and pleasantness are more accurately estimated. Thus, with all three parameters included, the conditional $R^2$-value increases from model M1 to M3, for the three subjective ratings considered: $R^2 = 0.54$ (M1), 0.76 (M2), 0.80 (M3); $R^2 = 0.60$ (M1), 0.83 (M2), 0.84 (M3); and $R^2 = 0.59$ (M1), 0.76 (M2), 0.78 (M3), for reported loudness, annoyance and pleasantness respectively.

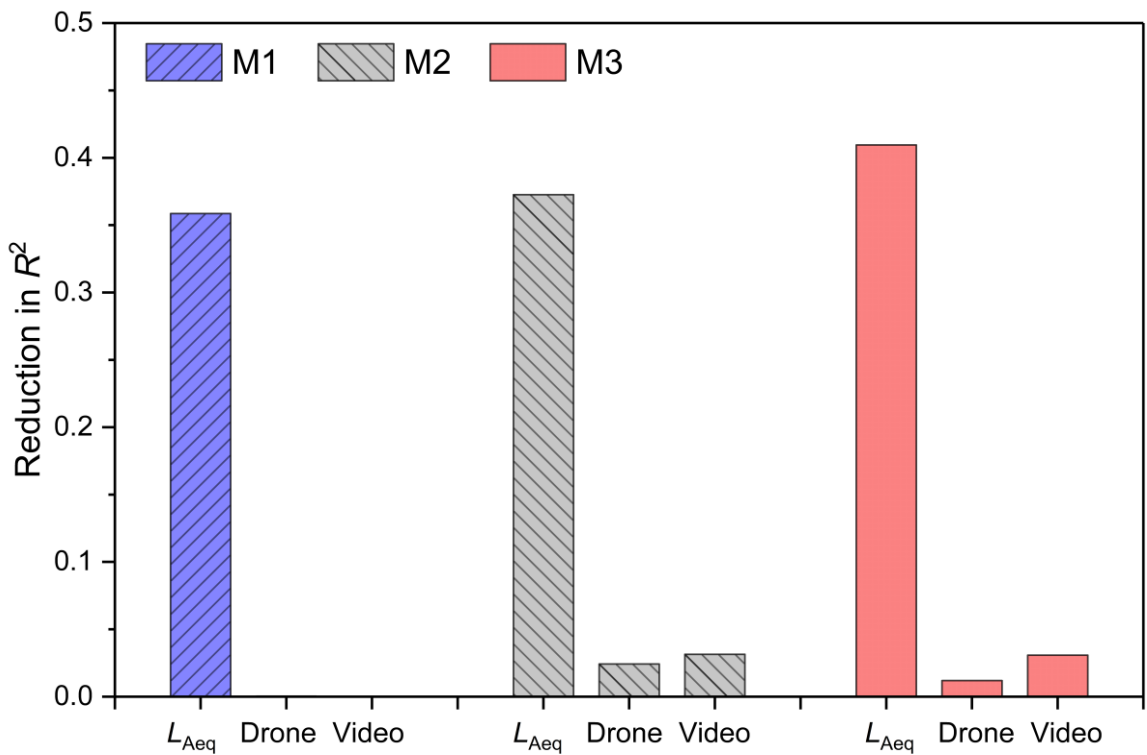

Figure 11. Reduction in conditional $R^2$ when subtracting $L_{Aeq}$, drone and video factors from the multilevel linear regression models M1 (fixed intercept, fixed slopes), M2 (fixed intercept, variable slopes) and M3 (variable intercept, variable slopes) for estimating the reported loudness.

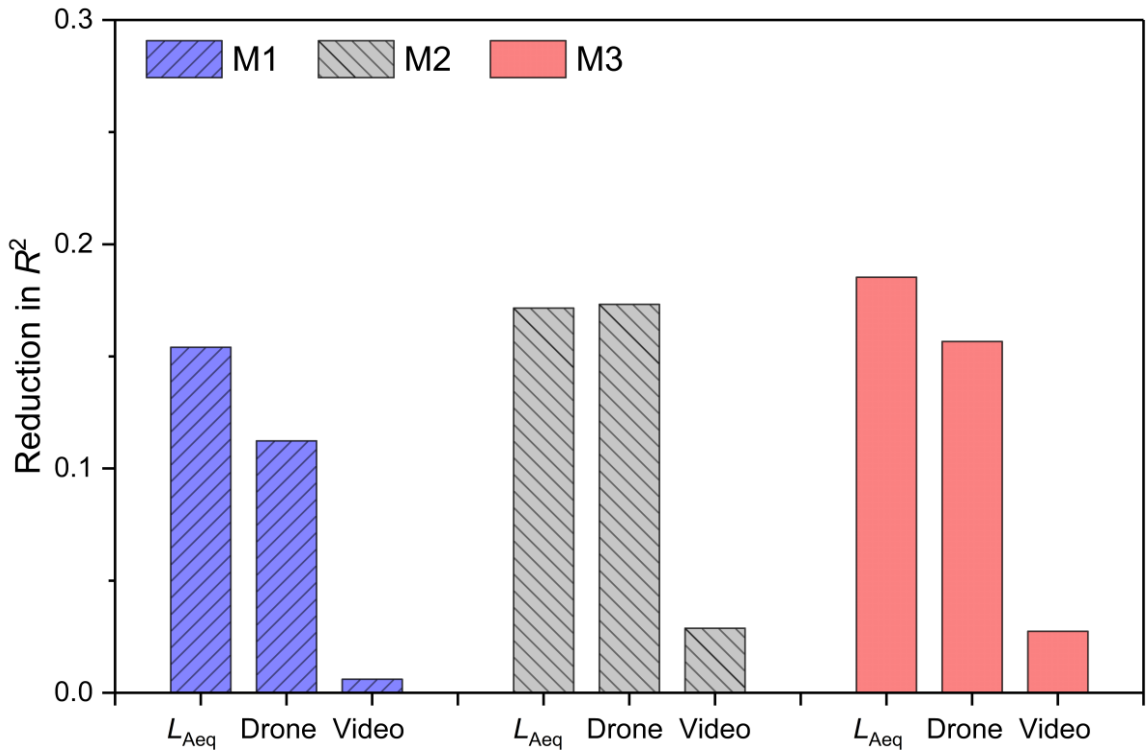

Figure 12. Reduction in conditional $R^2$ when subtracting $L_{Aeq}$, drone and video factors from the multilevel linear regression models M1 (fixed intercept, fixed slopes), M2 (fixed intercept, variable slopes) and M3 (variable intercept, variable slopes) for estimating the reported annoyance.

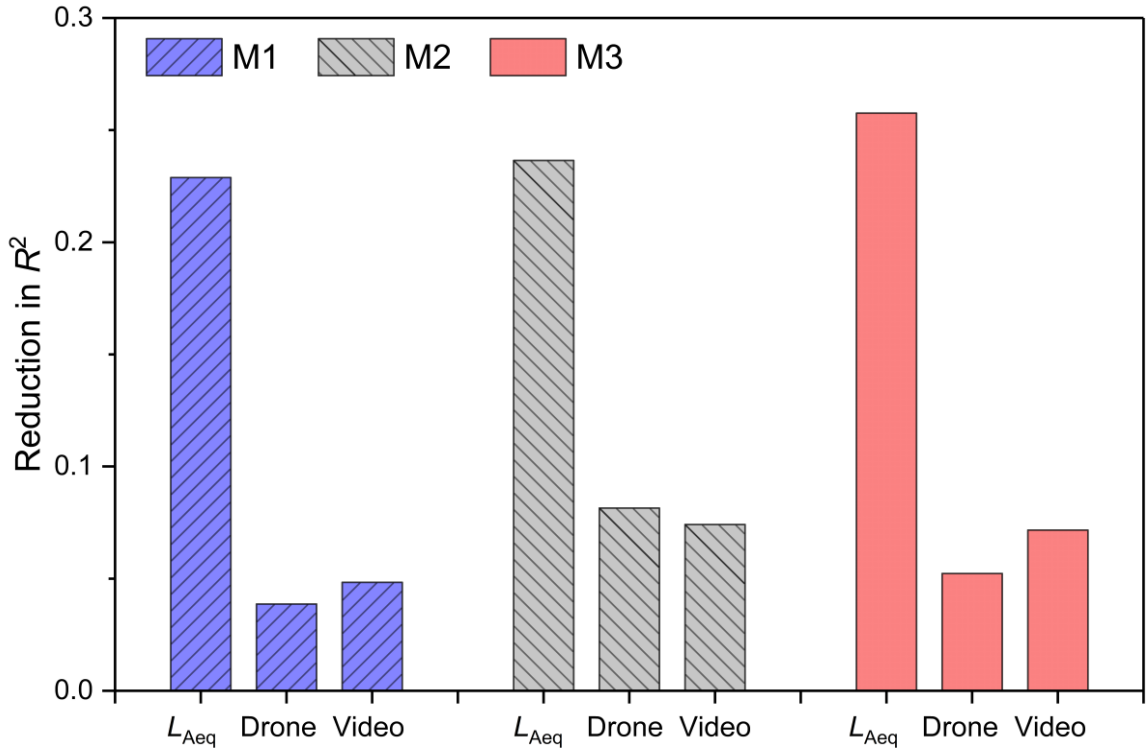

Figure 13. Reduction in conditional $R^2$ when subtracting $L_{\text{Aeq}}$, drone and video factors from the multilevel linear regression models M1 (fixed intercept, fixed slopes), M2 (fixed intercept, variable slopes) and M3 (variable intercept, variable slopes) for estimating the reported pleasantness.

As shown in Fig. 11, and in line with Fig. 9 – top, the estimation of the perceived loudness, as reported by the participants, is highly determined by $L_{\text{Aeq}}$ (reduction in $R^2$ between 0.36 and 0.41). The estimation of reported annoyance is equally determined by the factors $L_{\text{Aeq}}$ (reduction in $R^2$ between 0.15 and 0.19) and drone noise source (reduction in $R^2$ between 0.11 and 0.17) (Fig. 12). As described above (see Fig. 9 – middle), this finding confirms that participants' responses on perceived annoyance are also greatly influenced by acoustics (other than sound level) or non-acoustics factors associated to a small quadcopter noise source. Fig. 13 shows that $L_{\text{Aeq}}$ primarily determines the reported pleasantness (reduction in $R^2$ between 0.23 and 0.26). However, the factors drone noise source and, especially, visual stimuli (reduction in $R^2$ between 0.05 and 0.07) influence the participants' responses on perceived pleasantness.

## 4. Discussion

### 4.1. Influence of visual scenes on soundscape perception

Several authors (Hong et al., 2017; Puyana-Romero et al., 2017; Viollon et al., 2002) have confirmed the influence of visual scenes on soundscape perception. In the results presented in this paper (see Section 3.1), it is observed a decrease of the reported annoyance, in all urban scenarios tested, when visual stimuli is also presented. The use of visual stimuli leads also to a clear increase in the reported pleasantness, although statistically significant differences were only found in the noisiest locations (L1 and L4). In these locations, with high influence of road traffic noise, the visual scene modifies the soundscape perception towards an increase in perceived pleasantness (Pheasant et at., 2010). The human perception is multisensory by its very nature (Cassidy, 1997; Iachini et al., 2009; Pheasant et al., 2010), and therefore bi-modal stimuli (i.e. aural and visual) are essential for a full characterisation of soundscapes (Pheasant et al., 2010). Taking into account audio-visual interaction factors has been found to improve the reliability of studies evaluating the perception of soundscapes (Maffei et al., 2013, Ruotolo et al., 2013).

### 4.2. Combined effects of road traffic and drone noise

In locations with reduced influence of road traffic, statistically significant differences ($p < 0.05$) in reported loudness, annoyance and pleasantness are found between soundscapes with and without the noise of a small quadcopter hover (Table 2). In these locations, the presence of drone noise lead to significant increases in the reported annoyance and loudness, and significant decreases in reported pleasantness. Statistically significant differences in the perceived annoyance, reported by the participants, between soundscapes with and without drone noise are found in all locations tested. However, in the locations closest to road traffic (L1 and L4), the increase in reported annoyance with drone noise is very reduced, i.e. only about 1.3 times higher than without drone noise. In locations with little influence of road traffic noise (L2, L3, L5, L6 and L7), the reported annoyance with drone noise is up to 6.4 times higher than without drone noise.

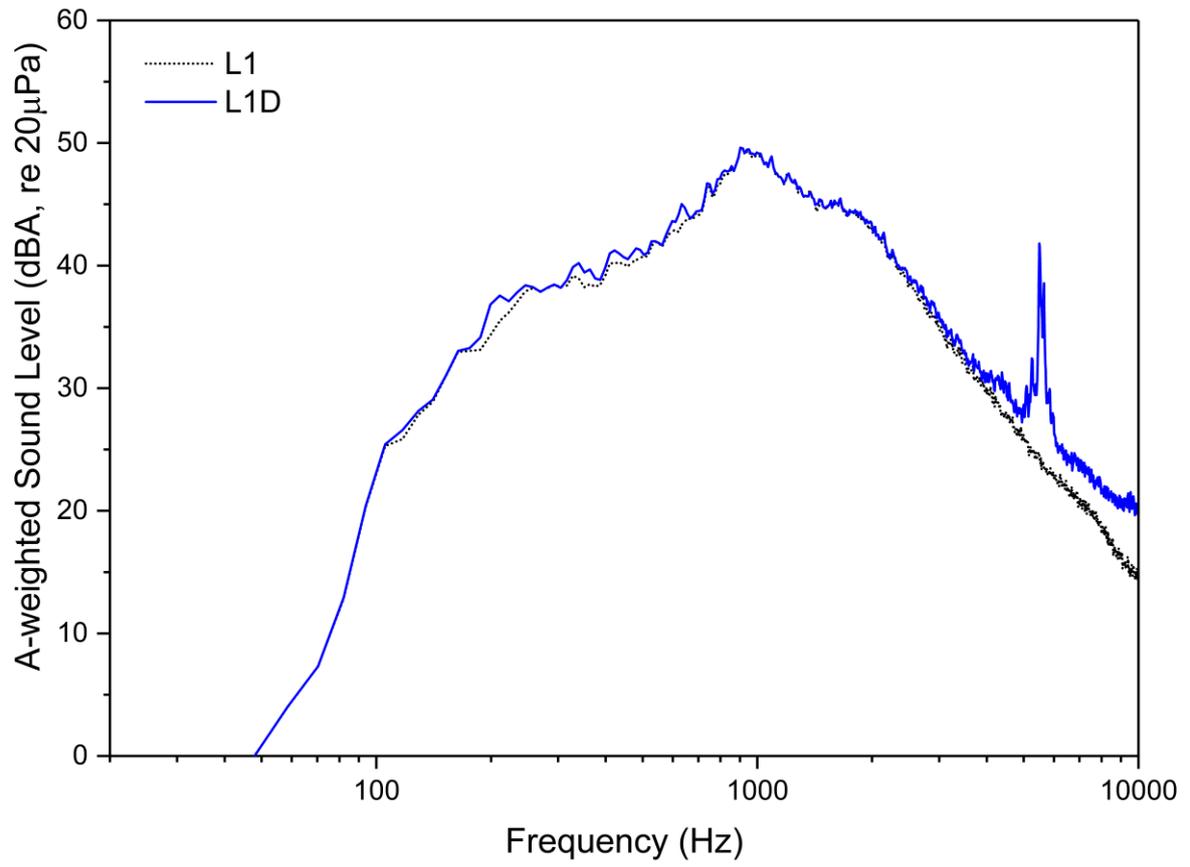

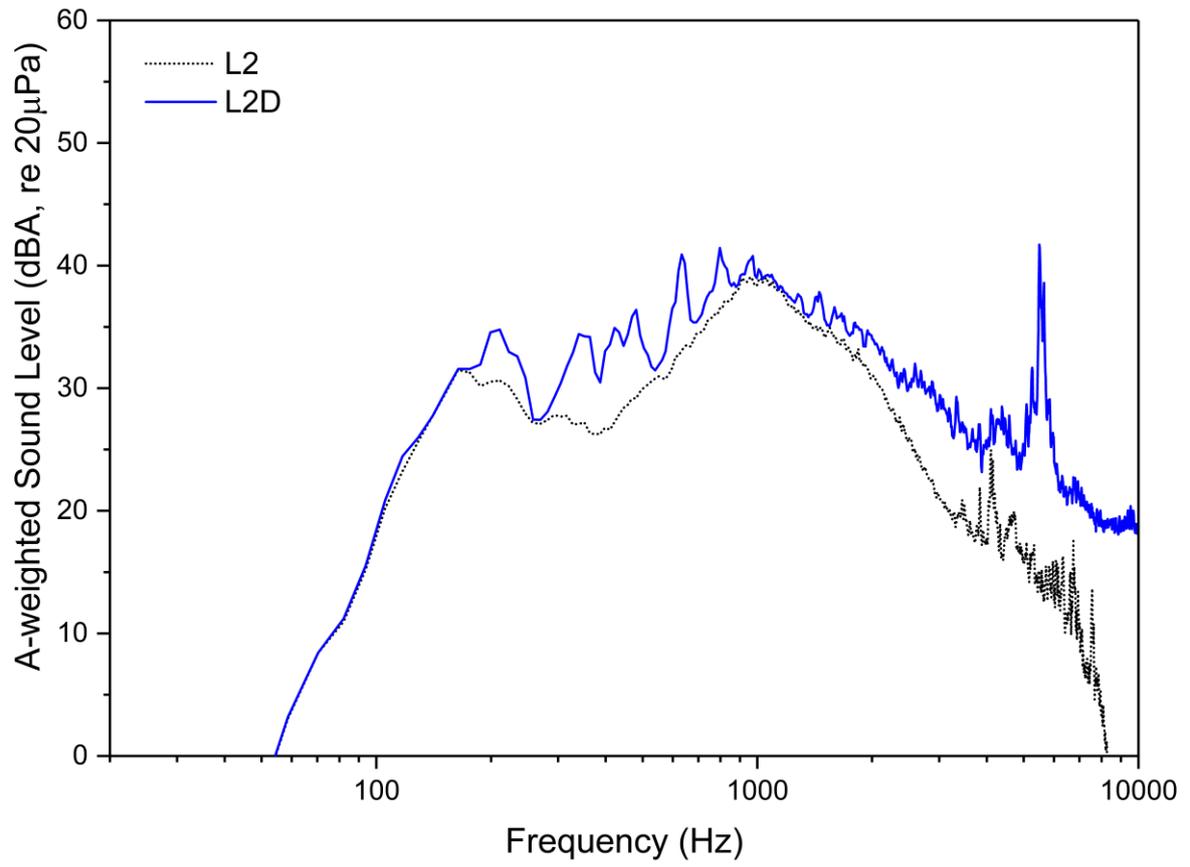

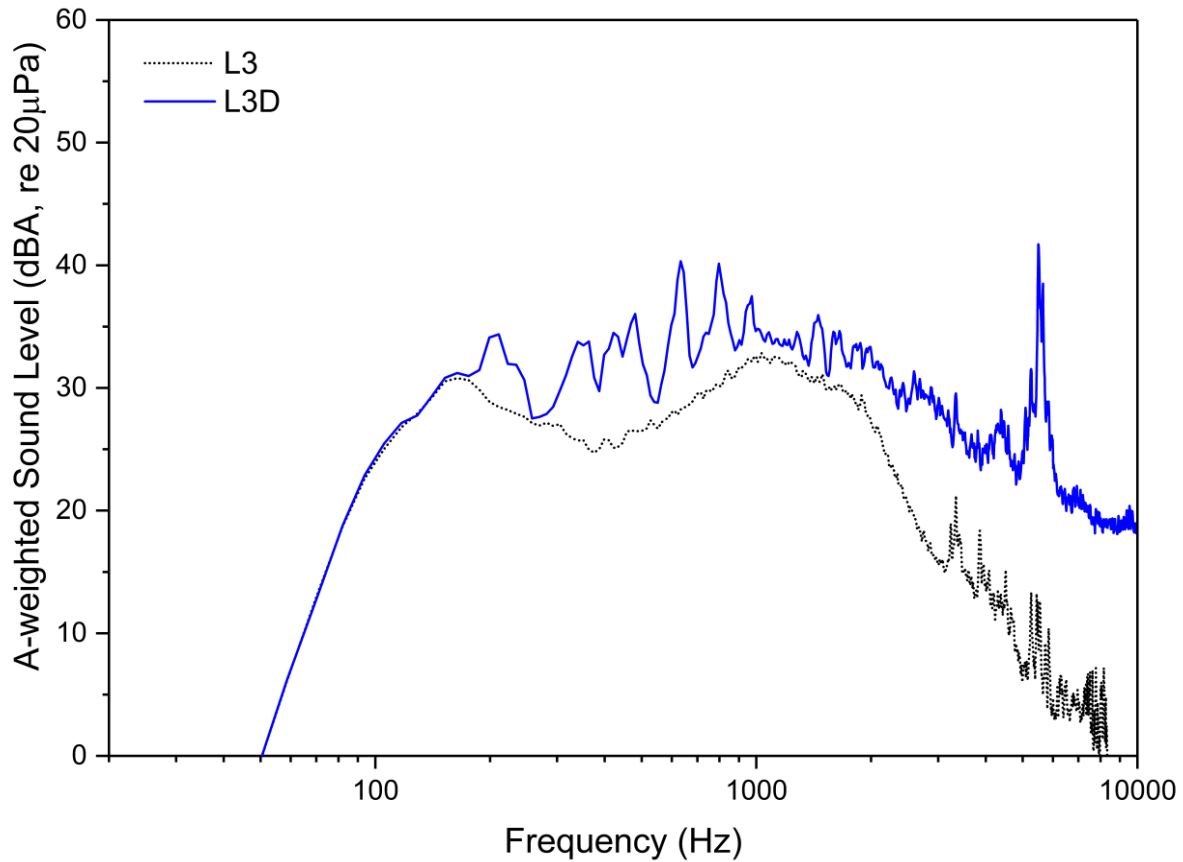

Figure 14. Frequency spectra (A-weighted Sound Pressure Level (dBA, re 20μPa)) measured in locations L1 (top), L2 (middle) and L3 (bottom), without (dotted line) and with (solid line) noise of the small quadcopter.

The overall sound level ($L_{Aeq}$) is the primary factor in determining the reported loudness for both soundscapes with and without drone noise (see Section 3.3). In determining reported annoyance for soundscapes with drone noise, the factor drone noise source is as important as $L_{Aeq}$ (see Fig. 12). In determining reported pleasantness for soundscapes with drone noise, $L_{Aeq}$ is the primary factor, but factor drone noise source, and especially visual factor influence the participants' responses. In Sections 3.2 and 3.3, it is hypothesised that the participants' responses on perceived annoyance and pleasantness for soundscapes with drone noise might be highly influenced by acoustics factors particularly characteristic of a small drone (quadcopter). The noise generated by a small quadcopter is mainly tonal in character, with a series of tones at harmonics of the blade passing frequency (BPF) of the rotors distributed across the frequency spectrum, and with a significant content in high frequency content consequence of the operation of the electric motors (Cabell et al., 2016; Torija et al., 2019b). Both the tonal and high frequency content are of significant importance for the subjective response to aircraft noise (Torija et. al, 2019a). Neither the tonality nor the very high frequency (above 4000 Hz) noise are taken into account in the $L_{Aeq}$ metric, which might be the reason of its poor performance in assessing the reported annoyance (and pleasantness) of soundscapes with drone noise (see Fig. 10). As shown in Fig. 14, in locations close to a road (Fig. 14 – top), the road traffic noise masks the noise generated by the small quadcopter, with the exception of the very high frequency noise. Under outdoor conditions, with flyovers at a particular altitude (e.g. 15-30 m and up to 100 m (Christian and Cabell, 2017)), the very high frequency noise is rapidly attenuated by atmospheric absorption. At locations further away from road traffic, with lower levels of road traffic noise, the tonal and high frequency content of the small quadcopter becomes more dominant (Fig. 14 – middle and bottom). Under these conditions, and assuming a linear relationship between the subjective ratings evaluated and $L_{Aeq}$, the participants' responses (on perceived annoyance and pleasantness) are mainly driven by the noise features of the small quadcopter, and are almost independent of the overall $L_{Aeq}$ in the location. In these locations, the perceived annoyance is reported as high as in locations with higher overall $L_{Aeq}$ (see Fig. 10 – middle).

These results suggest that, notwithstanding the potential safety issues, the development of corridors along busy roads for drone fleets to operate might reduce the overall community noise impact in urban areas. This will also avoid the disturbance of (urban) quiet areas (Iglesias-Merchan et al., 2015).

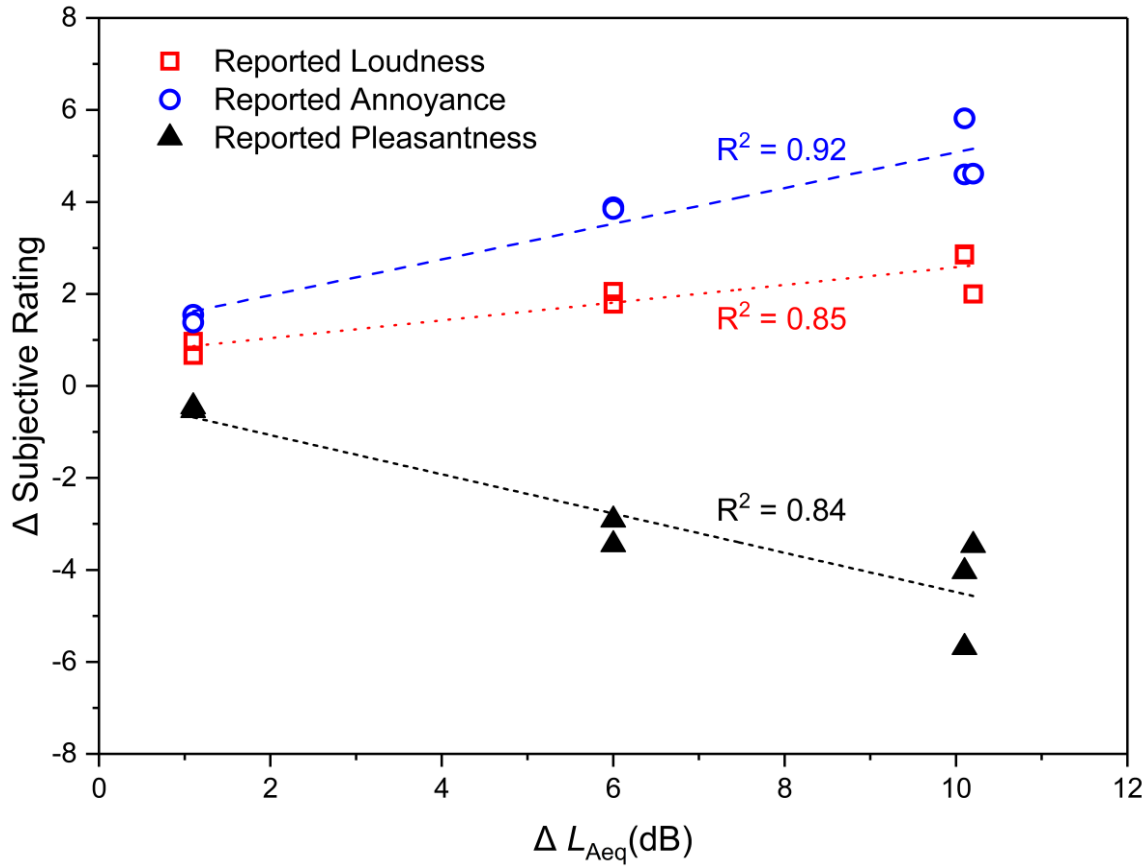

Figure 15. Changes in the subjective ratings loudness (squares), annoyance (circles) and pleasantness (triangles), and in the $L_{Aeq}$ without and with the noise generated by the drone hover, in the seven locations tested.

As seen in Fig. 15, the change in the reported loudness, annoyance and pleasantness between the soundscapes without and with drone noise is highly correlated with the increase of $L_{Aeq}$ generated by the small quadcopter over the ambient noise. Moreover, Fig. 15 shows that for all the locations tested, the increase in reported annoyance with drone noise is higher than the increase in reported loudness, which also suggests the influence of the tonal and high frequency content of drone noise (in addition to loudness) on the participants' responses.

In Sections 3.2 and 3.3, it is also hypothesised that the responses on perceived annoyance might be influenced by non-acoustics factors associated to the drone noise source. Although this research does not provide enough evidence to test this hypothesis, the participants' responses on perceived loudness and annoyance in location L7 (park without influence of road traffic, dominated by birds and water sounds) seem to suggest some influence of non-acoustics factors. Thus, in Fig. 15, the increase in reported annoyance and decrease in reported pleasantness with drone noise is notably higher and lesser, respectively, compared to the increase/decrease in locations with similar $\Delta L_{Aeq}$. In this location, there is probably an expectation of tranquility and relaxation, and the presence of drone noise is more penalised (Pheasant et al., 2008).

### 4.3. Constrains and limitations

The design of this research was carefully planned to investigate the perception of the same drone operation (a small quadcopter hover) on several urban soundscapes with a varying level of road traffic noise (and varying sound sources). The underlying hypothesis is that road traffic could mask drone noise, and thus corridors for drone fleets might be defined along road infrastructure to alleviate the noise impact of residents. A single drone was used in this research, a small quadcopter, whose size and characteristics resemble with drones currently under investigation for several applications from parcel delivery to surveillance. The focus of this research is the changes in sound level and frequency spectral when a drone operation is introduced in a typical urban soundscape. To simplify the achievement of this objective, a hover operation was selected, with the drone in a fixed position working at full power. Under these conditions, the influence of varying operational regimes, doppler effect and atmospheric absorption was avoided, and only the drone sound emission was assessed. As no drone movement was simulated, and the focus was on a steady positioned drone with other sources in the background, the experimenters decided to use a monophonic signal to present stimuli to the participants.

The findings of this paper refer to a drone hover with a steady frequency spectrum. Under flyover conditions, or with significant influence of atmospheric disturbances such as wind gusts, the flight control system varying rotor rotational speeds to maintain vehicle stability will create an unsteady acoustic signature (Cabell et al., 2016; Torija et al., 2019b). Furthermore, during the landing and take-off maneuvers, the changes in power setting and rotor rotational speeds will change sound directivity and frequency spectra. Both the unsteadiness of the acoustic signature and the changes in directivity and frequency spectra are likely to affect the audibility of the drone noise, and therefore, might alter the road traffic noise vs. drone noise combination effects described above.

Under the assumption of a linear relationship between the subjective ratings evaluated and $L_{Aeq}$, Fig. 10 suggests that the annoyance and pleasantness reported by the participants are mainly driven by the noise features of the small quadcopter. The comparison between drone noise and other transportation noise at the same sound level ($L_{Aeq}$) will provide further insight into the effects of the particular noise features of drones on sound perception.

After the main principles of the effects of drone noise are understood (as described in this paper), further investigation on the effects of drones operating in (a wider diversity of) urban environments on the perceived soundscape would require the simulation of flyovers (and take-off and landing maneuvers) to account for both emission and propagation factors. A wider range of drones would need to be assessed, accounting for differences in size, power, and configuration (fixed wing vs. multicopter). From the soundscape perception point of view, the use of spatial reproduction techniques (e.g. headphone-based First-Order-Ambisonic (FOA) tracked binaural or FOA 2D speaker arrays), would allow the immersion and plausibility of simulations with moving sources (Hong, et al., 2019; Lam, et al., 2019). As masking is a complex phenomenon influenced by not only sound levels and frequency, but also spatial cues (Cerwén et al., 2017), the use of spatial audio reproduction techniques would increase the fidelity of simulations with combined road traffic and drone noise sources, allowing a more refine evaluation of the masking capabilities of road traffic.

## 5.    Conclusions

This research represents a first approach to quantify the effect on urban soundscapes of introducing drone operations. The paper presents the results of a series of experiments aimed to investigate the effects of drone noise on a diversity of urban soundscapes. An audio-visual recording of a small quadcopter, recorded in an anechoic aeroacoustics laboratory, was added to audio-visual recordings taken in seven urban locations of different type. Both audio and audio plus panoramic video stimuli (using VR techniques) were presented to a series of participants, who were asked to report their perceived loudness, annoyance and pleasantness for each one. The soundscapes of the seven locations evaluated differed in the influence of road traffic noise. In locations close to busy roads, road traffic noise seems to mask the noise generated by the small quadcopter (with the exception of very high frequency noise). In these locations, the reported annoyance for the soundscapes with drone noise is only 1.3 times higher than without drone noise. In locations with little influence of road traffic noise, the specific characteristics of drone noise (i.e. series of tones at harmonics of rotors' BPF and high frequency noise) dominate the soundscape. In these locations, the participants reported a perceived annoyance with drone noise up to 6.4 times higher than without drone noise. In these locations with low influence of road traffic noise, the reported annoyance was about 7 (scale from 0 to 10) with drone noise, regardless the overall $L_{Aeq}$ in the location. These results have two main implications: (1) The annoyance reported for the soundscape with the drone present was highly influenced by the particular characteristics of drone noise. The descriptor $L_{Aeq}$ does not account for the particular noise features of drone noise, so novel metrics will be required for providing an effective assessment of drone noise impact in urban settings. (2) Notwithstanding any potential safety issue, the operation of drone fleets through corridors along busy roads might significantly mitigate the increase of community noise impact caused.

The use of panoramic video had little influence on the responses on perceived loudness. However, the reported annoyance and pleasantness of the soundscapes tested with panoramic visual stimuli were notably different than with only audio stimuli. As previous studies suggest, the simulation of audio-visual scenes can aid a more accurate assessment of the noise impact of transportation systems on urban soundscapes.

The results presented in this paper should be taken with caution, as only one quadcopter model in a fixed position is assessed. This single drone noise condition was enough for the purposes of this paper, as the emphasis was to assess the noise impact of the same drone noise in different urban soundscapes, with varying influence of road traffic. However, in future research, a variety of flyover maneuvers (with different airspeed and altitude) of a wider range of drones will be investigated for a more comprehensive analysis of drone noise impact on urban areas. Further work will investigate different conditions with visual cues, where the drone is visible, partly visible and not visible, also taking into account different distances (i.e. flyover altitudes).

**Acknowledgements**

Dr Zhengguang Li would like to thank the China Scholarship Council for the funding received (ref: 201808330656). The authors would also like to thank Dr J.LT. Lawrence for the collaboration and advice on setting up measurements in the Anechoic Doak Laboratory at the ISVR.

**Supplementary material**

The data (including audio and panoramic visual stimuli) used for this research will be provided by the authors upon request.